\documentclass[hyper,notoc,12pt,letterpaper]{JHEP3} 

\usepackage{latexsym,amsmath,amsfonts,amssymb}

\numberwithin{equation}{section}

\newcommand{\Nugual}[1]{\ensuremath{\mathcal{N}= #1}}

\newcommand{\rep}[1]{\ensuremath{\mathbf{#1}}}

\numberwithin{equation}{section}

\newcommand{\be}{\begin{equation}} \newcommand{\ee}{\end{equation}}
\newcommand{\bea}{\begin{equation} \begin{aligned}} \newcommand{\eea}{\end{aligned} \end{equation}}

\newcommand{\cN}{\mathcal{N}}

\newcommand{\bC}{\mathbb{C}}

\newcommand{\bR}{\mathbb{R}}

\DeclareMathOperator{\Tr}{tr}
\DeclareMathOperator{\ch}{ch}

\def\cN{\mathcal{N}}

\def\bR{\mathbb{R}}
\def\bC{\mathbb{C}}
\def\SU{\mathrm{SU}}

\def\SO{\mathrm{SO}}

\def\U{\mathrm{U}}

\def\ch{\mathop{\mathrm{ch}}}

\title{\hbox{Liouville/Toda central charges from M5-branes}}

\let\AA\diamondsuit
\let\BB\heartsuit
\author{Luis F. Alday$^\BB$, Francesco Benini$^\AA$, Yuji Tachikawa$^\BB$\\

{\tt alday,yujitach@ias.edu, fbenini@princeton.edu}\\
$^\BB$ School of Natural Sciences, Institute for Advanced Study, \\
Princeton, NJ 08540, USA \\
$^\AA$ Department of Physics, Princeton University, \\
Princeton, NJ 08544, USA
}

\preprint{PUTP-2314}

\abstract{ We show that the central charge of the Liouville and ADE Toda theories
can be reproduced by equivariantly integrating the anomaly eight-form
of the corresponding six-dimensional $\cN=(0,2)$ theories, which
describe the low-energy dynamics of M5-branes.
}

\begin{document}
\section{Introduction}
$\cN=2$ supersymmetric field theories in four dimensions are very rich, both
from the physical and mathematical points of view.
Recently, it was observed in  \cite{Gaiotto:2009we} that many
$\cN=2$ theories can be understood in a unified manner by realizing them
as a compactification of six-dimensional $\cN=(0,2)$ theories on a Riemann surface.
Furthermore, it was noted in \cite{Alday:2009aq} that Nekrasov's partition function \cite{Nekrasov:2002qd}
of such theories (with $\SU(2)$ gauge groups) computes the conformal blocks of the Virasoro algebra.
It was also noted that the partition function on $S^4$, as given by \cite{Pestun:2007rz}, coincides
with the corresponding correlation function of the Liouville theory.
Soon this 2d-4d correspondence was extended in \cite{Wyllard:2009hg,Mironov:2009by} to the case of
$\SU(N)$ gauge groups where the Liouville theory generalizes to the $A_{N-1}$ Toda theory.\footnote{Note that the Liouville theory is equivalent to the $A_1$ Toda theory.}

Given that these 4d theories are engineered from theories on M5-branes, one would like to
understand the above correspondence in terms of string/M-theory. A step in this direction
was made in \cite{Dijkgraaf:2009pc,Bonelli:2009zp}. Hinted at by the results of \cite{Wyllard:2009hg} and \cite{Benini:2009mz}, in \cite{Bonelli:2009zp}
 an interesting observation was made,
namely that  the anomaly eight-form of the 6d $\cN=(0,2)$ theory of type $A_{N-1}$
and the central charge of the Toda theory of the same type have  similar structures: \begin{align}
I_8[A_{N-1}]&= (N-1) I_8(1) + N(N^2-1) \frac{p_2(N)}{24} \;, \label{1.1}\\
c_\text{Toda}[A_{N-1}]&= (N-1) + N(N^2-1) Q^2 \;.\label{1.2}
\end{align}

In this short note,  we show that \eqref{1.2} with the correct value for $Q$, namely $Q=(\epsilon_1+\epsilon_2)^2/(\epsilon_1\epsilon_2)$,
arises  from \eqref{1.1} if we consider the compactification
of the 6d $(0,2)$ theory on $\bR^4$ with equivariant parameters $\epsilon_{1,2}$.
Furthermore, we will see that this relation works for arbitrary theories of type $A,D$ and $E$.

\section{Computation}

The anomaly eight-form of one M5-brane \cite{Witten:1996hc} is \begin{equation}
I_8(1)=\frac{1}{48}\left[ p_2(NW)-p_2(TW)+\frac14\bigl(p_1(TW)-p_1(NW)\bigr)^2\right] \;,
\end{equation}
where $NW$ and $TW$ stand for the normal and the tangent bundles of the worldvolume $W$, respectively and $p_k$ denotes the $k$-th Pontryagin class.
Using this, the anomaly of  the $\cN=(0,2)$ theory of
type $G$ ($G=A_n,D_n,E_n$) can be written as \cite{Harvey:1998bx,Intriligator:2000eq,Yi:2001bz}\footnote{For $E$-type $\cN=(0,2)$ theory, this formula is only conjectural
and there has been no independent check, to our knowledge. We assume
the correctness of the formula.}
\begin{equation}
I_8[G]=r_G I_8(1) + d_G h_G \frac{p_2(NW)}{24}.\label{anomaly}
\end{equation}
Here $r_G$, $d_G$ and $h_G$ are the rank, the dimension, and the  Coxeter number
of the Lie algebra of type $G$, respectively. They are tabulated in Table~\ref{data}.
\begin{table}
\[\begin{array}{c|ccc}
G& r_G & d_G & h_G \\
\hline
A_{N-1} &  N-1 & N^2-1 & N \\
D_N & N & N(2N-1) & 2N-2 \\
E_6 & 6 & 78 & 12 \\
E_7 & 7 & 133& 18 \\
E_8 & 8 & 248& 30
\end{array}\]
\caption{Data of the Lie algebras of type $A$, $D$, $ E$.  Note that $r_G(h_G+1)=d_G$. \label{data}}
\end{table}

Now, we wrap the $(0,2)$-theory of type $G$ on a four-manifold $X_4$.
The 11d theory lives on \begin{equation}
\Sigma\times X_4   \times \bR^5,
\end{equation}  where $\Sigma$ is the worldsheet of the resulting 2d theory.
We take $X_4$ to be Euclidean and $\Sigma$ to be Lorentzian.
The supercharges decompose as:
\be
\rep{4}_+ \times \rep{4} \;\to\; \Big( \frac12, 2, 1, 2, \frac12 \Big) + \Big( \frac12, 2, 1, 2, -\frac12 \Big) + \Big( -\frac12, 1, 2, 2, \frac12 \Big) + \Big( -\frac12, 1, 2, 2, -\frac12 \Big) \;,
\ee
where we listed the representation contents under the decomposition \begin{equation}
\SO(5,1) \times \SO(5) \;\to\; \SO(1,1) \times  \SU(2)_l \times \SU(2)_r \times \SO(3) \times \SO(2) \;.
\end{equation}
Here we have decomposed $\SO(4)\simeq \SU(2)_l \times \SU(2)_r$
and $\SO(5)\supset \SO(3) \times \SO(2)$ .
The symplectic Majorana condition acts on each factor separately.

Let us twist $\bR^5$ over $X_4$ so that a fraction of the supersymmetry remains.
We embed the spin connection of the $\SU(2)_r$ factor into the $\SO(3)$ factor, that is
\be
\SU(2)_r \;\to\; \text{diagonal part of }\big[ \SU(2)_r \times \SO(3) \big] \;. \label{twisting}
\ee
Note that
the $\SO(3)$ factor is the standard $\SU(2)_R$ symmetry of the four-dimensional
theory  if we think of the setup as the compactification of the six-dimensional
theory on $\Sigma$, giving an $\cN=2$ theory on $X_4$. Therefore this
twist is the one used by \cite{Witten:1988ze}.

After the twist, we get the symmetry group $\SO(1,1) \times \SU(2)_l \times \SU(2)_r \times \SO(2)$
and supercharges
\be
\Big( \frac12, 2, 2, \frac12 \Big) + \Big( \frac12, 2, 2,{} -\frac12 \Big)
+
\Big( {}-\frac12, 1,1+3,  \frac12 \Big) + \Big({}- \frac12,1, 1+3,{} -\frac12 \Big)
\;.
\ee
The preserved supercharges (scalars under $\SU(2)_l\times \SU(2)_r$)
 form a two-dimensional \Nugual{(0,2)} superalgebra, with $\U(1)$ R-symmetry.\footnote{This twist is different from
the one obtained by wrapping M5-branes on a holomorphic 4-cycle
in a Calabi-Yau threefold \cite{Maldacena:1997de}.}

Let us exploit this 2d $\cN=(0,2)$ superalgebra.
We take the right-movers to be
the supersymmetric side.
It is known that the anomaly polynomial and the central charges are related via
\begin{equation}
I_4 = \frac{c_R}{6} c_1(F)^2 + \frac{c_L-c_R}{24} p_1(T\Sigma) \label{central},
\end{equation} where $F$ is the external $\U(1)$ bundle which couples to the
$\U(1)_R$ symmetry. Let us check this formula against free multiplets.
The anomaly polynomial of
a right-moving complex Weyl fermion with charge $q$ is  \begin{equation}
I_4=\ch(qF)\hat A(T\Sigma) \Bigm|_{4} = \frac{q^2}{2} c_1(F)^2 -\frac{p_1(T\Sigma)}{24} \;.
\end{equation} The right-moving chiral multiplet has one complex boson, whose anomaly
is the same as that of two neutral Weyl fermions
and one Weyl fermion with charge $1$. In total,
$I_4=c_1(F)^2/2-p_1(T\Sigma)/8$ with $(c_L,c_R)=(0,3)$. On the other hand, the left-moving free real boson has
$I_4=p_1(T\Sigma)/24$ with $(c_L,c_R)=(1,0)$. Both cases agree with \eqref{central}.

Now let us determine $I_4$ of the compactified theory by integrating $I_8$ over $X_4$.
Let us assign the Chern roots as follows: $\pm t$ for the tangent bundle of $\Sigma$;
$\pm\lambda_1$, $\pm\lambda_2$ for the tangent bundle of $X_4$;  and
$\pm n_1$, $\pm n_2$, 0 for the normal bundle.  We include the $\U(1)$ R-symmetry through
\be
n_1 \;\to\;  2c_1(F) \;,
\ee
and the twisting \eqref{twisting} introduces
\be
n_2 \;\to\; \lambda_1 + \lambda_2 \;.\label{ttt}
\ee
Note that the doublet of $\SU(2)_r$ has
the Chern roots $\pm(\lambda_1+\lambda_2)/2$.
$(n_2,0,-n_2)$ should then be the Chern roots of the triplet, resulting in  \eqref{ttt}.

Then we evaluate the anomaly polynomial. Notice that $\lambda_1$ and $\lambda_2$ will be integrated over $X_4$. Since the 2d spacetime effectively behaves as four dimensional inside the anomaly polynomial,
forms whose degree along $T\Sigma$ is  higher than four automatically vanish. We get:
\begin{multline}
I_4 = \Big[ \frac{r_G + 2d_Gh_G}{12} \int \big( \lambda_1^2 + \lambda_2^2 \big) + \frac{(3r_G + 4d_Gh_G)}{12} \int \lambda_1\lambda_2 \Big] c_1(F)^2 \\
- \Big[ \frac{r_G}{48} \int \big( \lambda_1^2 + \lambda_2^2 \big) + \frac{r_G}{48} \int \lambda_1\lambda_2 \Big] p_1(T\Sigma) \;.
\end{multline}

Translating to $c_{L,R}$ using \eqref{central}, we find
\begin{equation}
\begin{aligned}
c_R&=\frac{1}{2} \big( P_1(X_4) +3 \chi(X_4) \big) r_G + \big( P_1(X_4)+2\chi(X_4) \big)d_Gh_G \;,\\
c_L&=\chi(X_4) r_G + \big( P_1(X_4)+2 \chi(X_4) \big) d_G h_G \;.
\end{aligned} \label{results}
\end{equation}
 Here, $\chi(X_4)=\int_{X_4} e(X_4) $ is the Euler number of $X_4$,
and  $P_1(X_4)=\int_{X_4} p_1(X_4)$
is the integrated first Pontryagin class which is three times  the signature of $X_4$.

For example,
let us wrap one M5-brane on $X_4=K3$, in which case there is effectively no twisting.
We start from $I_8(1)$ instead of $I_8[G]$, which effectively means using
$r_G=1$ and $d_Gh_G=0$ in \eqref{results}.
Using $P_1(K3)=-48$ and $\chi(K3)=24$, we obtain \begin{equation}
c_L=24,\qquad c_R=12
\end{equation} which is the value for the heterotic string, as it should be.

The case we are most interested in is
$X_4=\bR^4$, considering the
characteristic classes in the equivariant sense\footnote{Equivariant cohomology is a cohomology theory  which also captures the action of a group on a space. 
For simplicity we only consider the abelian case $\U(1)^n$.
Consider the space of differential forms on $M$ valued in the polynomial of the 
formal parameters $\epsilon_a$, ($a=1,\ldots,n$),  
and consider the deformed  differential $D_{\epsilon}=d+\epsilon_a \iota_{k^a}$. 
Here $\iota$ is the interior product and 
$k^a$ is the Killing vector of the $a$-th $\U(1)$.
Then $D_\epsilon{}^2 = \epsilon_a \pounds_{k_a}$ where $\pounds_{k_a}$ is the Lie derivative by $k_a$.  We define the equivariant cohomology $H_{\U(1)^n}(M)$ to be the cohomology of $D_\epsilon$ on the space of differential forms invariant under $\U(1)^n$.  Note that the formal parameters $\epsilon_a$ have degree $2$.
Equivariant characteristic classes are elements of the equivariant cohomology.
For example, consider $\bC$ acted on by $\U(1)$ which rotates the phase, and let the equivariant parameter be $\epsilon$. The  Chern class $c_1(T\bC)$ 
in the standard sense is of course trivial, 
but the equivariant Chern class is given by $c_1(T\bC)=\epsilon$. For more details, see e.g.~\cite{Libnei}.
}.
We take the action of $\U(1)^2$ to rotate two orthogonal two-planes in $\bR^4$,
and call the equivariant parameters $\epsilon_{1,2}$ respectively. 
The Chern classes of the two two-planes are $\epsilon_{1,2}$. Thus
 we have $p_1(T\bR^4)=\epsilon_1^2+\epsilon_2^2$ and $e(T\bR^4)=\epsilon_1\epsilon_2$. We then use the localization formula,
in the case where the fixed points are isolated: 
\[
\int_M \alpha = \sum_p \frac{\alpha|_p}{e(N_p)} \;.
\] 
The summation is over the fixed points $p$, and $e(N_p)$ is the equivariant Euler class
of the normal bundle of $p$ inside $M$.  In our case the only fixed point is the origin.
Therefore we have \begin{equation}
P_1(\bR^4)=\frac{\epsilon_1^2+\epsilon_2^2}{\epsilon_1 \epsilon_2},\qquad  \chi(\bR^4)=1.
\end{equation} Applying  \eqref{results},
we find
\begin{equation}
\begin{aligned}
c_R&=\frac{\epsilon_1^2+3\epsilon_1\epsilon_2 + \epsilon_2^2}{2\epsilon_1\epsilon_2} r_G + \frac{(\epsilon_1+\epsilon_2)^2}{\epsilon_1\epsilon_2} d_G h_G \;,\\
c_L&= r_G +  \frac{(\epsilon_1+\epsilon_2)^2}{\epsilon_1\epsilon_2} d_G h_G \;.
\end{aligned}
\end{equation}

Upon the identification $\epsilon_1/\epsilon_2=b^2$
advocated in \cite{Alday:2009aq},
$c_L$ perfectly agrees with
the central charge of the conformal Toda theory of type $G$ \cite{Hollowood:1989ep}:
\begin{equation}
c_\text{Toda}[G]= r_G +  \left(b+\frac1b\right)^2 d_G h_G \;.
\end{equation}

\section{Discussion}
A couple of comments are in order. First, recall that in the construction of \cite{Gaiotto:2009we} the $\cN=2$ theories are obtained by wrapping M5-branes
on $\bR^4\times \Sigma$, with a suitable twist on $\Sigma$ which preserves
one half of the supersymmetry.  So far, we have not taken  this twist into account.
When we perform it, the right-moving sector, which was
the supersymmetric part,  becomes topological and so $c_R\to 0$, while $c_L$
is untouched and agrees with the central charge of the Liouville/Toda theories.
This is consistent with the fact that Nekrasov's partition function
computes the chiral half of the Liouville/Toda correlation functions.

Second, notice that Nekrasov's partition function was computed
after introducing an equivariant deformation of $\bR^4$ by a
$\U(1)^2$ action with parameters $\epsilon_{1,2}$.
More precisely, the symmetry of the 4d theory is \begin{equation}
\SO(4)\times \SU(2)_R \simeq \SU(2)_l \times \SU(2)_r \times \SU(2)_R.
\end{equation} The topological theory has a modified Lorentz group
\begin{equation}
\SO(4)'\simeq \SU(2)_l \times \SU(2)_{r'} \;,
\end{equation} where  $\SU(2)_{r'}$ is
the diagonal subgroup of $\SU(2)_r\times \SU(2)_R$.
The $\U(1)^2$ used in the equivariant deformation
is the Cartan subgroup of this modified $\SO(4)'$.
This motivated our choice in \eqref{twisting}.
In view of this, it is also reasonable to evaluate
the anomaly polynomial in the same equivariant sense \footnote{Note that Nekrasov's partition function itself can be computed
as an equivariant integral over the instanton moduli space.}. It would be nice to have a better understanding of this point.

\section*{Acknowledgments}

It is a pleasure to thank G. Bonelli, J.~Maldacena, N.~Seiberg, A. Tanzini,
H.~Verlinde, B. Wecht and E.~Witten for helpful discussions.
LFA is supported in part by the DOE grant DE-FG02-90ER40542.
FB is supported by the DOE grant DE-FG02-91ER40671.
YT is supported in part by the NSF grant PHY-0503584, and by the Marvin L.
Goldberger membership at the Institute for Advanced Study.

\appendix
\section{Central charges of Sicilian gauge  theories of type $A$, $D$, $E$}

In \cite{Benini:2009mz} the central charges $a$ and $c$ of the 4d superconformal Sicilian theories of $A$ type (obtained by wrapping M5-branes on a genus-$g$ Riemann surface), both in the $\cN=2$ and $\cN=1$ case, were computed from the 6d anomaly polynomial. We observe that from \eqref{anomaly} the computation can be performed for the whole ADE series.

Let us start with the $\cN=2$ case.
Using the same Chern roots as in section 2, the line bundle of the \Nugual{1} R-symmetry is incorporated by: $n_1 \to n_1 + \frac23 c_1(F)$, $n_2 \to n_2 + \frac43 c_1(F)$. \Nugual{2} SUSY requires $n_1 + t = 0$, $n_2 = 0$. The integral over the Riemann surface is $\int_\Sigma t = 2 - 2g$.

The 4d 't Hooft anomalies of $U(1)_R$ are read from the formula:
\be
I_6 = \frac{\Tr R^3}{6} \, c_1(F)^3 - \frac{\Tr R}{24} \, c_1(F) p_1(T_4) \;.
\ee
Comparing this with the integral of $I_8$, we get:
\be
\Tr R^3 = \frac2{27} (g-1) (13 r_G + 16 d_G h_G) \qquad\qquad \Tr R = \frac23 (g-1) r_G \;.
\ee
Using the standard relations between $a$, $c$ and $\Tr R$, $\Tr R^3$, we get:
\be
a = (g-1) \frac{5r_G + 8 d_G h_G}{24} \qquad\qquad\qquad c = (g-1) \frac{r_G + 2 d_G h_G}{6} \;.
\ee
This agrees with \cite{Gaiotto:2009gz} for the $A$ series, and with \cite{Tachikawa:2009rb} for the $D$ series.
Similar formulas can be obtained in the $\cN=1$ case. The R-symmetry bundle
is given by $n_1 \to n_1 + c_1(F)$ and $n_2 \to n_2 + c_1(F)$, while $\cN=1$ SUSY
requires $n_1 + n_2 + t = 0$. We get:
\be
a = (g-1) \, \frac{6r_G + 9 d_Gh_G}{32} \qquad\qquad
c = (g-1) \, \frac{4r_G + 9 d_Gh_G}{32} \;.
\ee
%

\bibliographystyle{utphys}
\bibliography{bib}{}

\providecommand{\href}[2]{#2}\begingroup\raggedright\begin{thebibliography}{10}

\bibitem{Gaiotto:2009we}
D.~Gaiotto, ``{${\mathcal{N}}\!=2$ Dualities},''
\href{http://arxiv.org/abs/0904.2715}{{\tt arXiv:0904.2715 [hep-th]}}.

\bibitem{Alday:2009aq}
L.~F. Alday, D.~Gaiotto, and Y.~Tachikawa, ``{Liouville Correlation Functions
  from Four-Dimensional Gauge Theories},''
\href{http://arxiv.org/abs/0906.3219}{{\tt arXiv:0906.3219 [hep-th]}}.

\bibitem{Nekrasov:2002qd}
N.~A. Nekrasov, ``{Seiberg-Witten Prepotential from Instanton Counting},'' {\em
  Adv. Theor. Math. Phys.} {\bf 7} (2004)  831--864,
\href{http://arxiv.org/abs/hep-th/0206161}{{\tt arXiv:hep-th/0206161}}.

\bibitem{Pestun:2007rz}
V.~Pestun, ``{Localization of Gauge Theory on a Four-Sphere and Supersymmetric
  Wilson Loops},''
\href{http://arxiv.org/abs/0712.2824}{{\tt arXiv:0712.2824 [hep-th]}}.

\bibitem{Wyllard:2009hg}
N.~Wyllard, ``{$A_{N-1}$ Conformal Toda Field Theory Correlation Functions from
  Conformal ${\mathcal{N}}\!=2$ $SU(N)$ Quiver Gauge Theories},''
\href{http://arxiv.org/abs/0907.2189}{{\tt arXiv:0907.2189 [hep-th]}}.

\bibitem{Mironov:2009by}
A.~Mironov and A.~Morozov, ``{On AGT Relation in the Case of U(3)},''
\href{http://arxiv.org/abs/0908.2569}{{\tt arXiv:0908.2569 [hep-th]}}.

\bibitem{Dijkgraaf:2009pc}
R.~Dijkgraaf and C.~Vafa, ``{Toda Theories, Matrix Models, Topological Strings,
  and ${\mathcal{N}}\!=2$ Gauge Systems},''
\href{http://arxiv.org/abs/0909.2453}{{\tt arXiv:0909.2453 [hep-th]}}.

\bibitem{Bonelli:2009zp}
G.~Bonelli and A.~Tanzini, ``{Hitchin Systems, ${\mathcal{N}}\!=2$ Gauge
  Theories and W-Gravity},''
\href{http://arxiv.org/abs/0909.4031}{{\tt arXiv:0909.4031 [hep-th]}}.

\bibitem{Benini:2009mz}
F.~Benini, Y.~Tachikawa, and B.~Wecht, ``{Sicilian Gauge Theories and
  ${\mathcal{N}}\!=1$ Dualities},''
\href{http://arxiv.org/abs/0909.1327}{{\tt arXiv:0909.1327 [hep-th]}}.

\bibitem{Witten:1996hc}
E.~Witten, ``{Five-Brane Effective Action in M-Theory},''
  \href{http://dx.doi.org/10.1016/S0393-0440(97)80160-X}{{\em J. Geom. Phys.}
  {\bf 22} (1997)  103--133},
\href{http://arxiv.org/abs/hep-th/9610234}{{\tt arXiv:hep-th/9610234}}.

\bibitem{Harvey:1998bx}
J.~A. Harvey, R.~Minasian, and G.~W. Moore, ``{Non-Abelian Tensor-Multiplet
  Anomalies},'' {\em JHEP} {\bf 09} (1998)  004,
\href{http://arxiv.org/abs/hep-th/9808060}{{\tt arXiv:hep-th/9808060}}.

\bibitem{Intriligator:2000eq}
K.~A. Intriligator, ``{Anomaly Matching and a Hopf-Wess-Zumino Term in 6D, $N$
  = (2,0) Field Theories},''
  \href{http://dx.doi.org/10.1016/S0550-3213(00)00148-6}{{\em Nucl. Phys.} {\bf
  B581} (2000)  257--273},
\href{http://arxiv.org/abs/hep-th/0001205}{{\tt arXiv:hep-th/0001205}}.

\bibitem{Yi:2001bz}
P.~Yi, ``{Anomaly of (2,0) Theories},''
  \href{http://dx.doi.org/10.1103/PhysRevD.64.106006}{{\em Phys. Rev.} {\bf
  D64} (2001)  106006},
\href{http://arxiv.org/abs/hep-th/0106165}{{\tt arXiv:hep-th/0106165}}.

\bibitem{Witten:1988ze}
E.~Witten, ``{Topological Quantum Field Theory},''
\href{http://dx.doi.org/10.1007/BF01223371}{{\em Commun. Math. Phys.} {\bf 117}
  (1988)  353}.

\bibitem{Maldacena:1997de}
J.~M. Maldacena, A.~Strominger, and E.~Witten, ``{Black Hole Entropy in
  M-Theory},'' {\em JHEP} {\bf 12} (1997)  002,
\href{http://arxiv.org/abs/hep-th/9711053}{{\tt arXiv:hep-th/9711053}}.

\bibitem{Libnei}
M.~Libnei, ``Lecture notes on equivariant cohomology,''
  \href{http://arxiv.org/abs/0709.3615}{{\tt arXiv:0709.3615 [math]}}.

\bibitem{Hollowood:1989ep}
T.~J. Hollowood and P.~Mansfield, ``{Quantum Group Structure of Quantum Toda
  Conformal Field Theories. 1},''
\href{http://dx.doi.org/10.1016/0550-3213(90)90129-2}{{\em Nucl. Phys.} {\bf
  B330} (1990)  720}.

\bibitem{Gaiotto:2009gz}
D.~Gaiotto and J.~Maldacena, ``{The Gravity Duals of ${\mathcal{N}}\!=2$
  Superconformal Field Theories},''
\href{http://arxiv.org/abs/0904.4466}{{\tt arXiv:0904.4466 [hep-th]}}.

\bibitem{Tachikawa:2009rb}
Y.~Tachikawa, ``{Six-Dimensional $D_{N}$ Theory and Four-Dimensional SO-USp
  Quivers},''
\href{http://arxiv.org/abs/0905.4074}{{\tt arXiv:0905.4074 [hep-th]}}.

\end{thebibliography}\endgroup

\end{document}